\documentclass[manuscript]{aastex}
\usepackage{apjfonts}
\usepackage{natbib}

\shorttitle{Helioseismic Constraints on the Solar Ne/O}
\shortauthors{Delahaye et al.}

\begin{document}

\title{Helioseismic Constraints on the Solar Ne/O Ratio and Heavy Element Abundances.}

\author{F. Delahaye \altaffilmark{1}, M. H. Pinsonneault \altaffilmark{2}, L. Pinsonneault \altaffilmark{3} and C. J. Zeippen  \altaffilmark{1} }
\affil{1:LERMA, Observatoire de Paris, ENS, UPMC, UCP, CNRS ; 5 Place Jules Janssen, 92190 Meudon, France \\
2: Astronomy Dept., The Ohio State University, 140W 18th ave., 43210 Columbus OH \\
3: 6859 Langford Drive, Edina, MN  55436}

\begin{abstract}

We examine the constraints imposed by helioseismic data on the solar heavy element abundances. In prior work we argued that the measured depth of the surface convection zone $R_{CZ}$ and the surface helium abundance $Y_{surf} $ were good metallicity indicators which placed separable constraints on light metals (CNONe) and the heavier species with good relative meteoritic abundances. The resulting interiors-based abundance scale was higher than some published studies based on 3D model atmospheres at a highly significant level.  In this paper we explore the usage of the solar sound speed in the radiative interior as an additional diagnostic, and find that it is sensitive to changes in the Ne/O ratio even for models constructed to have the same 
$R_{CZ}$ and $Y_{surf}$. Three distinct helioseismic tests (opacity in the radiative core, ionization in the convection zone, and the core mean molecular weight) yield consistent results. Our preferred O, Ne and Fe abundances are $8.86\pm0.04$, $8.15\pm0.17$ and $7.50\pm0.05$ respectively. They are consistent with the midrange of recently published 3D atmospheric abundances measurements. 
The values for O, Ne and Fe which combine interiors and atmospheric inferences are $8.83\pm0.04 $, $8.08\pm0.09 $ and $7.49\pm0.04 $ respectively. 

\end{abstract}

\keywords{atomic data - opacity - Sun: abundances - Sun: evolution - Sun: interior}

\section{Introduction}

The Sun is a fundamental benchmark in astronomical abundance studies. Meteorites can be used to reliably infer relative abundances for numerous heavy elements and we have detailed information about limb darkening, precise fundamental parameters, and a wide variety
of diagnostic features.  A decade ago the problem of solar abundances seemed settled. A series of papers \citep[][GS98 hereafter]{ag89, gn93, gs98} had established a consistent set of surface abundances, and interiors models \citep[e.g.][]{Bahcall2001} were in good agreement with helioseismic data ranging from sound speed as a function of depth to surface helium and convection zone depth measurements.  However, the atmospheres models used in this generation of study were one-dimensional and had important simplifying assumptions about the physics of the atmospheres.  With the advent of a new generation of numerical simulations it became possible to relax these severe assumptions.  For many heavier elements such as iron the revisions in abundance were small and within the errors.  However, the lighter elements (CNO) were a different story.  These species have problematic photospheric abundance diagnostics under solar conditions and meteorites cannot be used to infer their primordial relative abundGances. The 3D model atmospheres, along with other revisions such as enhanced non-LTE effects and the identification of blending features in commonly used abundance indicators,  led to much lower CNO abundances \citep[][AGS05 hereafter]{ags05}.  The revised abundances placed solar interiors models in stark disagreement with helioseismic data \citep[see, for example][]{AB05,BP04,BBPS05, Guzik05, Montalban04, Turck04}. This solar abundance problem spurred a re-examination of the opacities \citep{Seaton04, Badnell05}, which improved opacity data as compared to the original source from the Opacity Project \citep{SYMP1994, CJZ1995} but did not resolve the problem.

The next generation of helioseismic studies took a different approach, inverting the problem and using distinct characteristics to infer a higher seismic abundance more in line with the GS98 scale \citep[][DP06 hereafter]{AB07, Chaplin07,DP06}. DP06 identified the depth of the surface convection zone and the surface helium abundance as sensitive abundance diagnostics.  This data could be used to predict absolute light and heavy metal abundances when combined with relative meteoritic abundances for heavy metals and relative photospheric/coronal abundances for CNONe. \citet{AB07} used the depression of the adiabatic gradient induced by metal ionization, mostly from oxygen, to infer the absolute metallicty of the solar convection zone and obtained a comparably high abundance. \citet{Chaplin07} derived a similar result from modelling the metallicity dependance of the mean molecular weight in the deep solar core.

There was a subsequent burst of revisions in the atmospheric abundances. The initial generation of models were criticised for problems in fitting the observed solar limb-darkening \citep{ayres06, Koesterke08}.  Substantial revisions of
the original model atmospheres led to a signficant improvement in the fit to solar data and removed the mild tension between the abundances of heavier elements (Fe, Si) and seismic inferences \citep[][hereafter AGSS09]{AGSS09}.  However, the estimates for CNO abundances remained quite low.  Interestingly, a different group of investigators with an independant 3D atmospheres code had comparable agreement with solar data but found higher photospheric abundances \citep{caffau08, caffau10}. The solar abundance problem therefore takes on a different 
cast; the distinction is not between new 3D and old 1D atmospheres, but rather on differing choices of abundance indicators and different ways of analyzing them.  A similar point was made by us in a prior paper \citep{PD09}, where a reanalysis of the AGS05 work found an atmospheric abundance scale intermediate between seismology and the prior 3D values, comparable to the one obtained by \citet{caffau10}.

However, neon has always been an important, and underconstrained, ingredient in the puzzle. Ne is a significant opacity source in stellar interiors, so the helioseismic problems induced by low oxygen could be compensated for if the solar Ne was sufficiently high \citep{AB05,BBS05}. There are also chemical evolution arguments in favor of such a model \citep{Turck04}. Ne is not preserved in meteorites in significant amounts and lacks photospheric diagnostics because of its atomic structure.  
As a result, indirect arguments from coronal 
abundance ratios must be used, and significant variations from the standard mixture have been argued for \citep{drake05,drake07}.
However, the bulk of evidence in the Sun appears to favor a lower ratio \citep{Schmelz05, Young05}; see AGSS09 for a detailed discussion.  

In DP06 we used only the scalar constraints ($R_{CZ} $ and $Y_{surf}$) as diagnostic features to infer abundances.  However, Ne retains electrons to higher temperatures than O does, and as a result different Ne+O mixtures with the same opacity at the base of the surface convection zone will have different opacity in deeper layers.  This leads to a distinct signature in the sound speed profile.  In this letter we update the mixture constraints based on the same method, but with improved stellar interiors model physics. By adding the sound speed profile as an additional constraint we can set interesting bounds on the solar $Ne/O$ ratio, and these are reinforced by considering the seismic metallicity bounds based on ionization and mean molecular weight.  
In section 2 we recall the method used in DP06, to determine the appropriate composition in order to reproduce the two observable scalars $R_{CZ}$ and $Y_{surf}$ and the extension of this method including the additional constraint $C_{sound}$ to determine 
the ratio $Ne/O$. In Section 3 we discuss our results.   

\section{Method}

In DP06 we presented our method to infer the solar heavy and light metal abundances required for theoretical models that reproduce the observed solar $Y_{surf} $ and $R_{CZ} $.  There are three essential steps involved in determining a seismic solar composition and its associated error.  First, the observed solar properties and their errors must be obtained.  Second, errors in the input solar model physics introduce uncertainties in the theoretically predicted $Y_{surf} $ and $R_{CZ} $. These uncertainties can be correlated; for example, increasing the degree of gravitational settling both deepens the model $R_{CZ} $ and decreases the $Y_{surf} $.  Both the difference between theory and observation and the associated error are now defined for a given solar composition.  
 
Changes in the abundance of the heavier metals primarily affect the $Y_{surf} $ through their impact on the central opacity and initial helium, while changes in the CNO abundances primarily affect $R_{CZ} $ . As a result, we can solve for the heavy (relative abundances inferred from meteorites) and light (relative abundances inferred from photospheric and coronal data) abundances consistent with the solar data as a function of the assumed $Ne/O$. We derived the seismic solar abundances in term of $A(Fe) $ and  $A(O)$ keeping the ratios $C/O$, $N/O$, $Ne/O$ and $X_i/Fe$ (where $X_i$ stands for all metals but CNONe) constant. $A(X)$ is the abundance of element $X$ on the logarithmic scale with $A(H)=12$. 
We obtained $A(O)_{AGS05}=8.87~dex \pm 0.041$ and $A(Fe)_{AGS05}=7.51~dex \pm 0.045$ for the calibrated model using the relative CNONe abundances from the AGS05 compilation. Similar results were obtained using relative light element abundances from GS98. 

For the present work we first updated our solar model inputs, using an equation of state table \citep{eos2006} with revised online data (http://adg.llnl.gov/Research/OPAL/opal.html, hereafter EOS2006), low temperature opacity tables from \citet{Ferguson2005} and updated nuclear reaction rates \citep{s17, s34b, s34a, s114}.  We then solve for the uncertainties in sound speed profile for models with fixed $R_{CZ} $ and $Y_{surf}$.  This defines a range of sound speed deviations which cannot be distinguished from theoretical errors, and which
therefore set a bound on our ability to infer the Ne/O ratio.

\subsection{ Uncertainty envelope for the sound speed profile}

The uncertainties associated with the sound speed profile ($c_{sound}$) include the observational errors in $c_{sound}^{Helio}$, uncertainties in the predicted $c_{sound}$ due to the uncertainties in $R_{CZ} $ and $Y_{surf}$ and finally the theoretical errors from uncertainties in the model input physics (opacities, equation of state, nuclear reaction cross sections, rate of gravitational settling) and assumed solar properties (age, radius, and luminosity). In the top panel of Figure 1 we illustrate individual theoretical uncertainties with lines and the total effect by the gray shade. On the bottom panel the measured $c_{sound}$ uncertainties are in light gray, the ones induced by observational uncertainties in $R_{CZ} $ and $Y_{surf}$ are in medium light gray and the theoretical error envelope is in medium dark gray. The total error envelope is represented in dark gray in the bottom panel. Our detailed procedure is described below.    

In order to estimate the effect of the different physics ingredients onto the $c_{sound}$ profile, we changed one at a time and adjusted the composition in the same manner as in DP06 to recover the seismic $R_{CZ} $ and $Y_{surf}$.  
The differences in $c_{sound}$ between these calibrated models are then measures of the impact of theoretical uncertainties on the $c_{sound}$ profile. 
We adopt the same error estimates and central values as described in Table I of DP06 with the following exceptions. The errors and central values of the nuclear reaction cross sections have been taken from the sources cited above and for the equation of state the errors are derived from the differences in the prediction between the EOS2001 and EOS2006 tables. 
We used the differences in the calibrated model sound speed profiles to infer the envelope of uncertainties. From the top panel of Figure 1, we see the uncertainties in the diffusion coefficients for gravitational settling (dot-dashed line) are the most significant error source, followed by the uncertainties in the opacities (solid line).

The observational uncertainties in the solar $c_{sound}$ are deduced from the range of inverted solar structures in \citet{Basu2000} (see Figure 1, bottom panel, light gray envelope). Errors in $R_{CZ} $ and $Y_{surf}$ also impact the $c_{sound}$ profile. The fractional depth of the convection zone is $R_{CZ}=0.7133 \pm 0.0005\ R_{\sun}$ \citep{BA2004}.  The theoretical
location of the convection zone depth can change by up to 0.001 with different opacity interpolation methods \citep{BSP04}; we therefore treat the latter as the uncertainty in $R_{CZ}$. The helium abundance in the surface convection zone ($Y_{surf}=0.2483 \pm 0.0046$)  is measured from the strength of the depression in the adiabatic gradient in the helium ionization zone \citep[][see DP06 for discussion]{Basu1998,Richard1998,DiMauro2002}. 
The observational errors in $R_{CZ}$ and $Y_{surf}$ permit a range of derived compositions which will then generates different 
predicted $c_{sound}$ profiles. The values $R_{CZ}^{Helio}$ and $Y_{surf}^{Helio}$ define a reference mixture. The maximum 
$c_{sound}$ profile change occurs for $R_{CZ} ^{\prime} = R_{CZ}^{Helio} + \sigma_{R_{CZ}^{Helio}}$ and $Y_{surf} ^{\prime} = Y_{surf}^{Helio} + \sigma_{Y_{surf}^{Helio}}$. This is our one $\sigma$ errors envelope (the medium light gray shade in Figure 1, bottom panel). We added the three sources defined above in quadrature to obtain the total envelope for uncertainties in the sound speed profile difference (dark grey shade of Figure 1 bottom panel). The theoretical errors predominate.  

\section{Constraining $Ne/O$}

We are now in a position to extract additional information from the sound speed profile above and beyond those constraints imposed by the measurement of $R_{CZ} $ and $Y_{surf}$. Carbon and nitrogen are much less abundant than oxygen and have similar ionization structures. As a result, our method does not place interesting constraints on the $C/O$ and $N/O$ ratios. However changes in Ne/O have a significant impact on the $c_{sound}$ profile, as seen in the top panel of Figure 2 where we compare different calibrated models to one reference model which has $(O,Ne,Fe,Ne/O)=(8.80,8.25,7.49,0.28)$. In the bottom panel we present the percentage difference between the solar $c_{sound}$ and the different calibrated models with A(O) ranging from 8.90 (top) to 8.70 (bottom) in 0.05 dex increaments. The dashed line correspond to our prefered model which minimizes the rms in the percentage difference in $c_{sound}$. Ne remains opaque to higher temperatures than oxygen, so a combination of higher Ne and lower O which yields the same opacity at the base of the convection zone will be more opaque deeper in the star. This difference is at a maximum relative to the theoretical uncertainties at 0.56 solar radii. Changes of $A(Ne)$ of $0.17$ dex at this radius are comparable to the 1 $\sigma$ total error. We treat 0.17 dex as our uncertainty on the $Ne/O$ ratio from the $c_{sound}$ profile independent of the scalar or coronal abundance constraints.

In Figure 3 we present the predicted $R_{CZ}$ and $Y_{surf}$ (top panel) and the percentage difference in $c_{sound}$ between the Sun and severals models using various compositions. As previously noted, low O scales are strongly disfavored in both the scalar constraints (top panel) and the $c_{sound}$ profile (bottom panel). All of the models are discrepant at a statistically significant level in the $c_{sound}$ immediately below the surface convection zone. We attribute this to the treatment of the mixing which is required to explain the low solar Li \citep{Pinsonneault1997}. As in DP06, we reduced the settling coefficient to account for the effect of rotational mixing from full evolutionary models \citep{Richard1996, Bahcall2001}. However rotational mixing requires a more complete model so we don't use this feature as a diagnostic and defer such a topic to a subsequend paper.
Also illustrated in the bottom panel of Figure 3 (dashed line) is a model with low O (value from AGSS09) and high Ne (to satisfy the helioseismic contraints $R_{CZ}$ and $Y_{surf}$) which is excluded at more than two $\sigma$.

\section{Conclusion}
  
Helioseismology has proven to be a powerful means of inferring the solar composition.  The depth of the surface convection zone and the initial helium abundance are very sensitive to the opacity of light and heavy metal respectively, and the errors in these model properties can be reliably quantified.  Reproducing these seismic scalar features require a relatively high solar metallicity.  If we combine these scalar constraints with reasonable priors about the relative abundances of CNONe (from the photosphere) and heavy elements (from meteorites) we can therefore infer a seismic solar mixture based primarily on stellar interiors physics. Recent revisions to the input physics of the solar model - improvements in the nuclear reaction cross-sections, low temperature opacities, and equation of state - yield a reference A(O) = 8.86 and A(Fe) = 7.50 which is the same as DP06.

However, the linkage between the thermal structure and the composition is indirect, which can lead to ambiguities in the interpretation of the strong constraints imposed by the measured solar thermal structure.  Fortunately, there is additional information encoded in the solar sound speed profile and there are independent helioseismic diagnostics which yield very similar results.

It is common to use differences in sound speed between models and the Sun as a measure of agreement. However, many of the striking visual features in such difference plots are in fact manifestations of errors in the independently determined scalar features, especially the convection zone depth.  We therefore argue that extracting information from the sound speed profile first requires calibration to a common set of observables, which can be achieved with modest adjustments of the solar composition.  Sound speed variations at the 0.1 - 0.2 \% level arise from reasonable changes in the input physics. This sets a threshold below which deviations between the Sun and the models can be explained by known observational and theoretical sources.

It is possible to construct different relative mixtures of the light species CNONe which have similar opacities at the base of the convection zone, and thus similar scalar properties.  Changes in C and N are weakly constrained because their ionization structures are similar to that of oxygen and they are much less abundant (by the same token, even large variations in C and N cannot produce a low surface oxygen).  Neon, by contrast, is abundant enough that large variations could change the estimated oxygen, and uncertain enough that such variations cannot be dismissed out of hand.  We find that changes in the Ne/O level produce a strong differential sound speed signature at $R=0.56 R_{\sun}$ which is explained by Ne retaining electrons (and thus higher opacity) to higher temperatures than O does.  Although the resulting bounds are modest (0.08 dex in oxygen), there would be a clear signal from a very high Ne / very low O mixture which are clearly not present in the data.  

Our method is sensitive to opacity, and as a result the most likely systematic error source is in the theoretical quantum mechanical opacity calculations despite the very good agreement between independent theoretical calculation. Experiments are underway to check on this possibility and transition to an abundance scale primarily based on experiments.  However the same high abundances are inferred from two other independent helioseismic diagnostics - the level of metal ionization in the convection zone and the mean molecular weight of the solar core.  In a comprehensive review, \citet{BA08} found that increases in neon did not reduce the need for a high oxygen in the convection zone (c.f. their Figure 21).  One would therefore need to invoke multiple errors in distinct interiors and atmospheric input physics, with the same sign and magnitude, to produce an acceptable high Ne - low O mixture.  

Our best interiors estimate for composition is $A(O)=8.86\pm0.04 $, $A(Ne)=8.15\pm0.17 $ and $A(Fe)=7.50\pm0.05$. \citet{Lodders2009} combined recent atmospheres measurements for estimates of $A(O)=8.73\pm0.07 $, $A(Ne)=8.05\pm0.10 $ and $A(Fe)=7.45\pm0.08$; these two scales are consistent within the errors. A weighted mean of the two leads to $A(O)=8.83\pm0.04 $, $A(Ne)=8.08\pm0.09 $ and $A(Fe)=7.49\pm0.05$ which we contend is the most precise current estimate for abundances. 
In DP06 we derived relationships quantifying combinations of C, N, O, and Ne with the same $R_{CZ}$ and $Y_{surf}$. In the case of Ne, nonlinear effects can be induced when the change in the Ne/O ratio is too large and a better fitting relationship is 
$A(O)=  -7.52\times (Ne/O)^4 +9.03\times (Ne/O)^3 - 3.10\times (Ne/O)^2 -0.32\times (Ne/O) + 8.98$

Our abundance scale is consistent with some photospheric abundance measurements \citep{caffau10,PD09} but not the lower AGSS09 values.  This is not a conflict between modern and primitive atmospheres treatments, or between one- and three-dimensional studies, but rather reflects differences between competing atmospheres models and judgment calls on the choice of indicators, continuum levels, and the proper treatment of blending features.  We are therefore hopeful that helioseismology may provide an absolute abundance standard which can be used to discriminate between competing models and which can be used to calibrate the appropriate composition diagnostics for the next generation of stellar models.
\acknowledgments
FD would like to thank C. Stehl\'e for financial support and
A. Formicola for information on updated nuclear reaction
cross-sections.  MP would like to acknowledge support from DOE grant DE-FG52-09NA29580. MP is grateful to the Observatoire de Paris for their hospitality and support during a visit in April-June 2010.

\clearpage




\begin{figure}
\centerline{
\includegraphics[width=11.0cm,angle=0,clip=true]{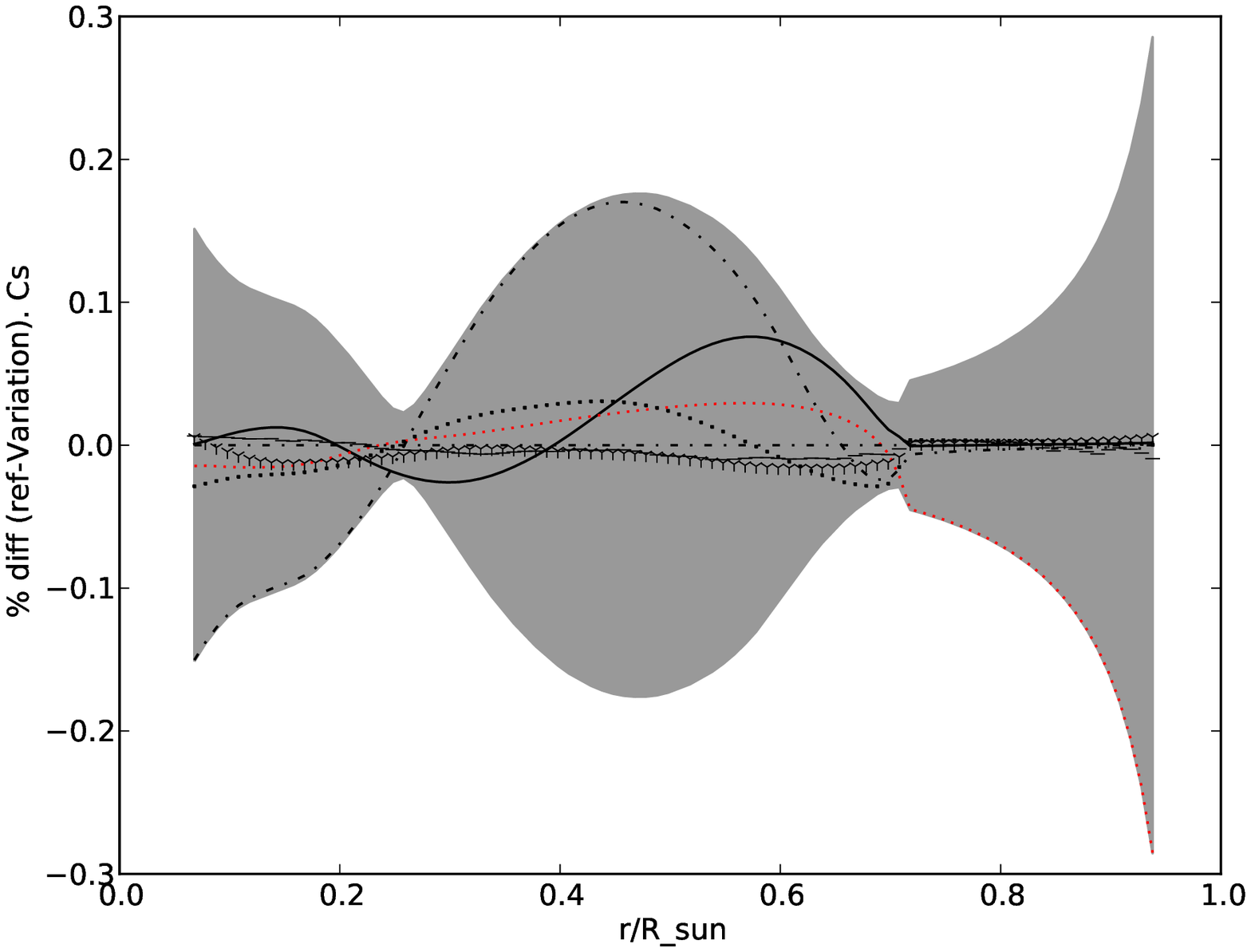}}  
\centerline{
\includegraphics[width=11.0cm,angle=0,clip=true]{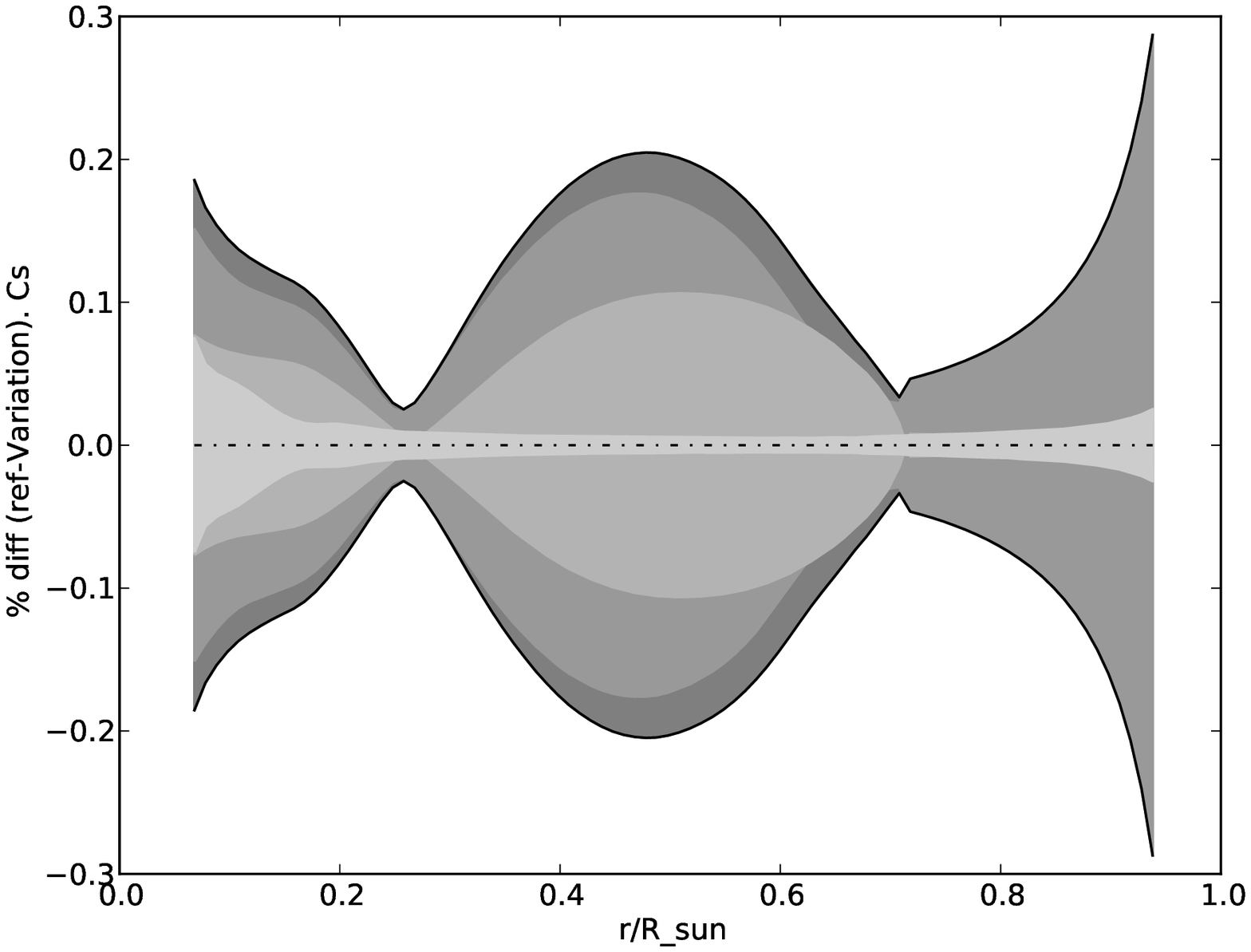}}
\caption{Top Panel: Effect of physics changes on the predicted sound speed profile for models with the composition adjusted to reproduce the same Rcz and Ysurf. Detailled Uncertainties: Dot-Dashed (Magenta) $\rightarrow$ Settling $\pm 16\% $ , Solid (Black) $\rightarrow \kappa _R (OPvsOPAL)$, dotted (Red) $\rightarrow \sigma (R_{\sun})$, square (Green) $\rightarrow$ Age $\pm 0.01 Gyr$, Y (Blue) $\rightarrow \ L_{\sun}\pm 4\%$, long dashed (Cyan) $\rightarrow $ EOS: Bottom panel: errors in the inverted solar sound speed profile (light gray), sound speed errors arising from uncertainties in the fitted Rcz and Ysurf (medium light gray), theoretical errors from the top panel (medium dark gray) and the total error budget (dark gray) from adding all of the above in quadrature.}
\end{figure}

\begin{figure}
\epsscale{0.5}
\centerline{
\includegraphics[width=12.0cm,angle=0,clip=true]{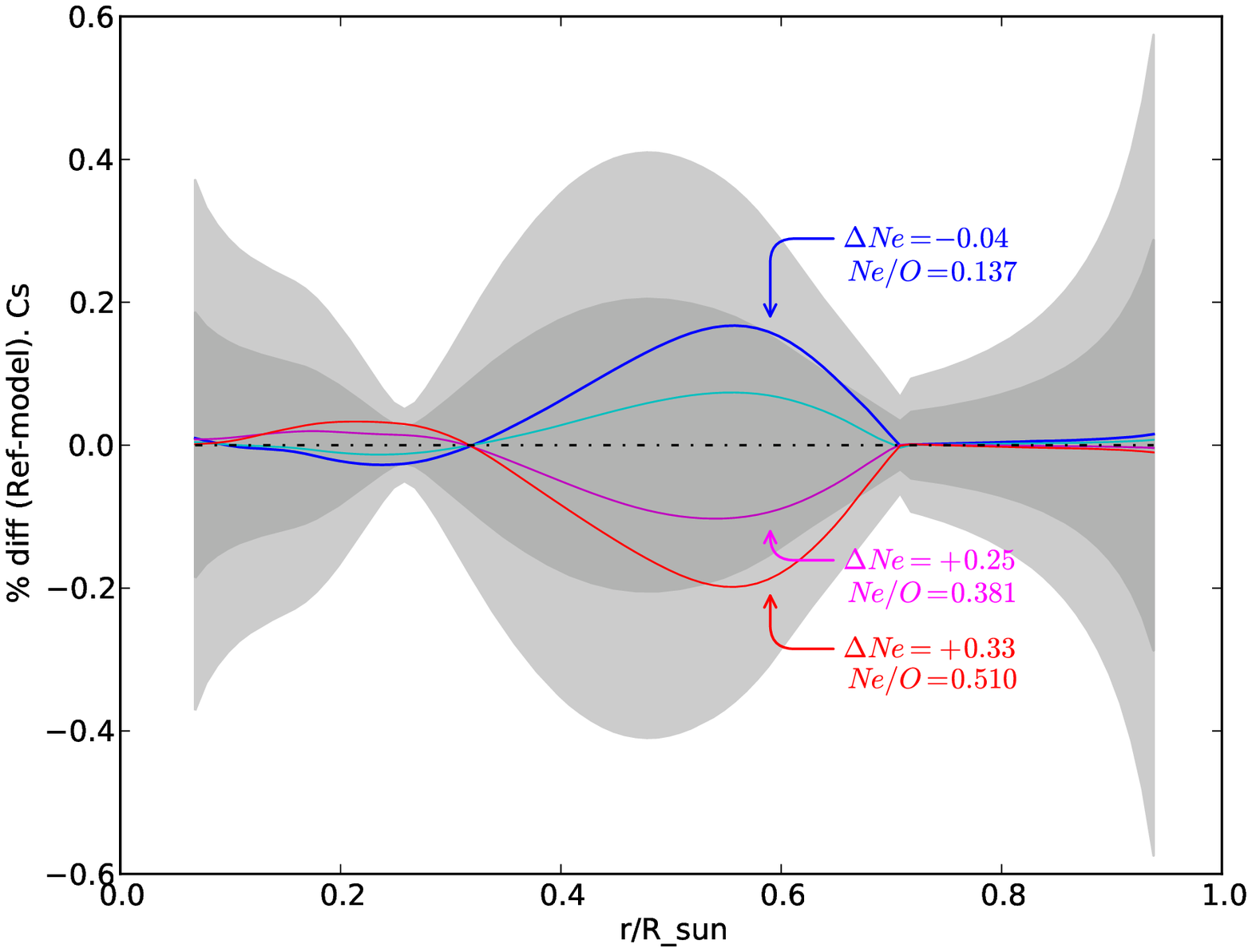}}    
\centerline{
\includegraphics[width=12.0cm,angle=0,clip=true]{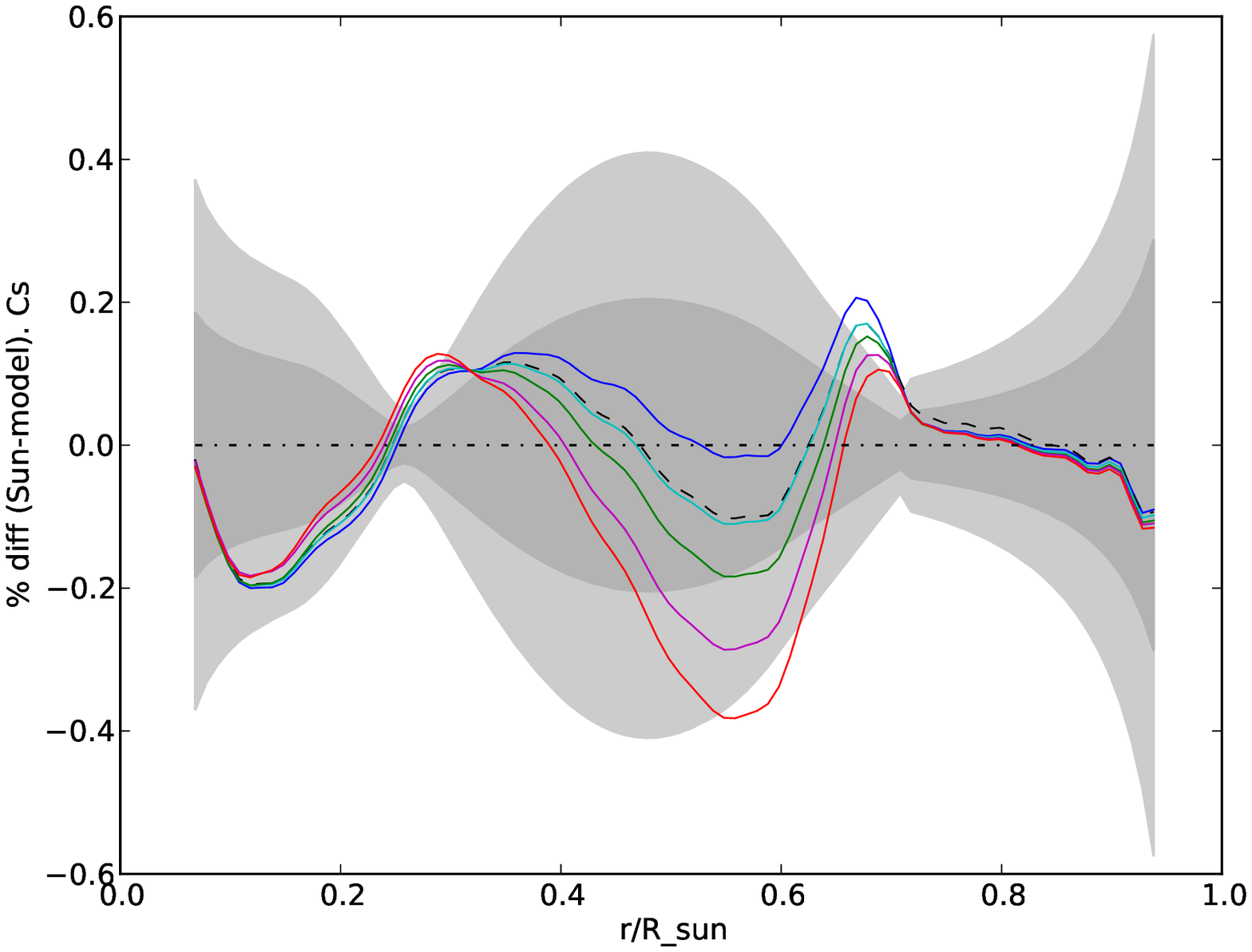}}
\caption{Models with different Ne/O ratio but the same $R_{CZ}$ and $Y_{surf}$ are compared. These models have Ne and Fe-group abundances adjusted to fit $R^{CZ}_{\sun}$ and $Y^{surf}_{\sun}$. Top Panel: They are compared to a reference model (see text). 
Bottom panel: The same models are compared to the Sun. The dashed line corresponds to our preferred model with $(O,Ne,Fe,Ne/O)=8.86,8.15,7.50,0.197)$ which minimises the rms for the sound speed. Cases with A(O) from 8.90 (top) to 8.70 (bottom) in 0.05 dex increments are also shown (solid lines).
\label{fig2}}
\end{figure}

\begin{figure}
\epsscale{1.0}
\centerline{
\includegraphics[scale=.60]{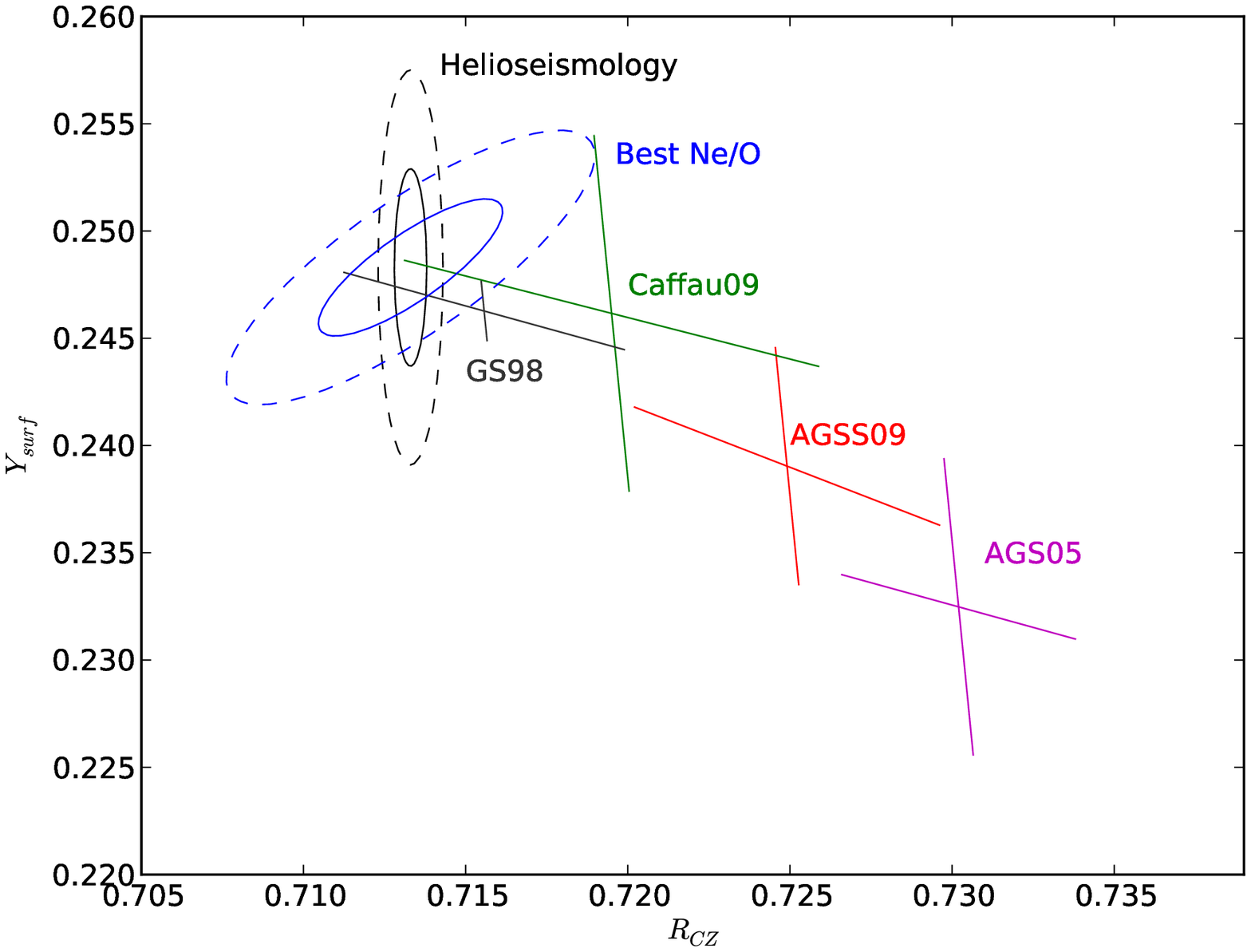}}
\centerline{\includegraphics[scale=.60]{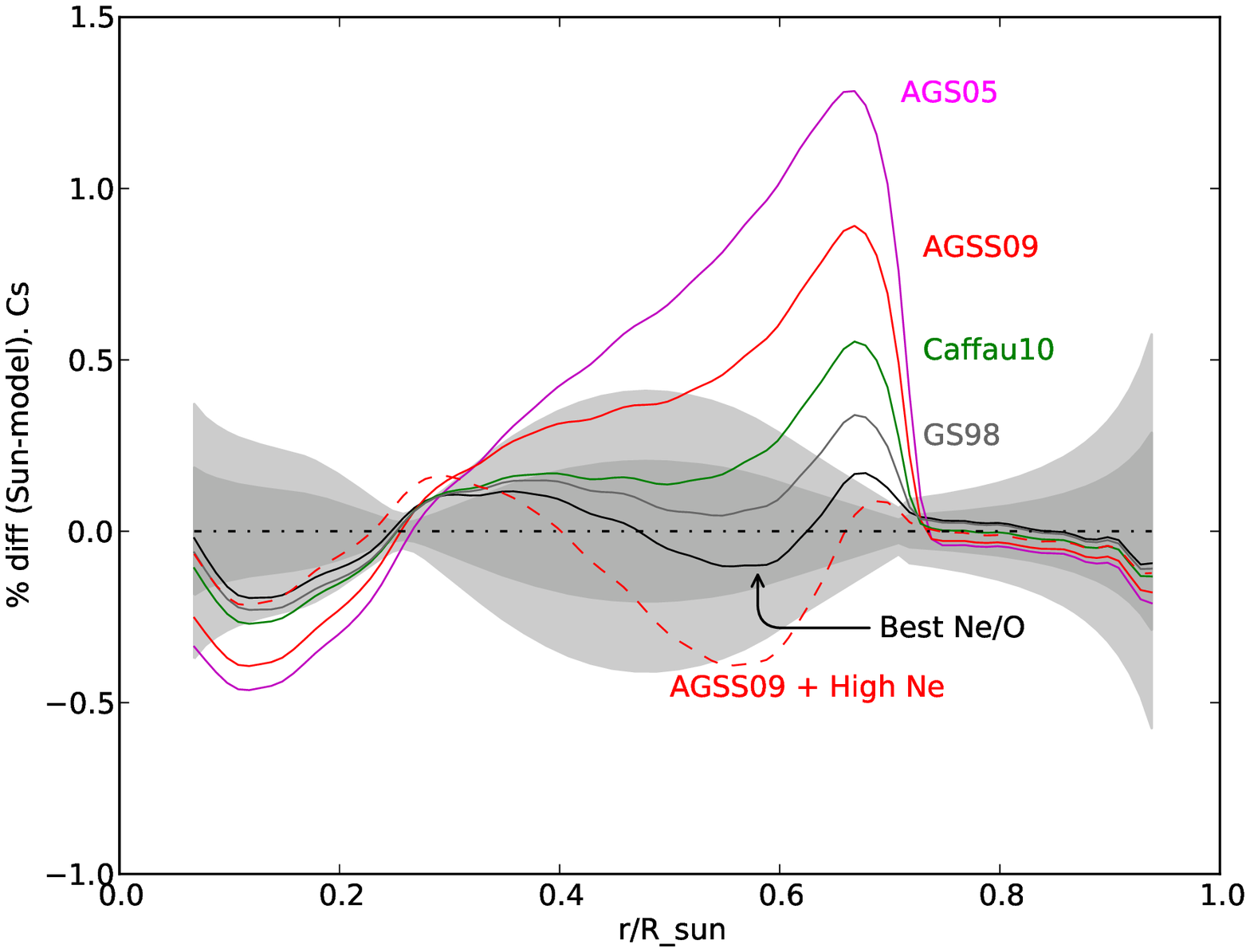}}
\caption{Published solar mixtures compared with the scalar Rcz and Ysurf constraints (top panel) and sound speed profile (bottom panel). All models have a settling coefficients reduced by 20\% to account for rotational mixing. In this figure AGS05 $\rightarrow$ Asplund et al. 2005, AGSS09 $\rightarrow$ Asplund et al. 2010, Caffau10 $\rightarrow$ Caffau et al. 2010, GS98 $\rightarrow$ Grevesse \& Sauval 1998 and Best Ne/O is a modified GS98 mixture with Fe-group and Ne adjusted for seismic constraints. 
\label{fig3}}
\end{figure}





\begin{thebibliography}{}

\bibitem[Anders \& Grevesse(1989)]{ag89} Anders, E., \& Grevesse, N.\ 1989, \gca, 53, 197 
\bibitem[Antia \& Basu (2005)]{AB05} Antia, M., \& Basu, S. 2005, \apjl, 620, L129 
\bibitem[Antia \& Basu(2007)]{AB07} Antia, H.~M., \& Basu, S.\ 2007, Astronomische Nachrichten, 328, 257 
\bibitem[Asplund et al.(2005)]{ags05} Asplund, M., Grevesse, N., \& Sauval, A. J. 2005, Cosmic Abundances as Records of Stellar Evolution and Nucleosynthesis, 336, 25
\bibitem[Asplund et al.(2009)]{AGSS09} Asplund, M., Grevesse, N., Sauval, A.~J., \& Scott, P.\ 2009, \araa, 47, 481 
\bibitem[Ayres et al.(2006)]{ayres06} Ayres, T.~R., Plymate, C., \& Keller, C.~U.\ 2006, \apjs, 165, 618 
\bibitem[Badnell et al.(2005)]{Badnell05} Badnell, N. R., Bautista, M. A., Butler, K., Delahaye, F., Mendoza, C., Palmeri, P., Zeippen, C. J., \& Seaton, M. J. 2005, \mnras, 360, 458 
\bibitem[Bahcall et al.(2005a)]{BBPS05} Bahcall, J. N., Basu, S., Pinsonneault, M. H., \& Serenelli 2005a, ApJ, 618, 1049
\bibitem[Bahcall et al.(2005b)]{BBS05} Bahcall, J. N., Basu, S. \& Serenelli, A.M. 2005b, \apj, 631, 1281
\bibitem[Bahcall et al.(2005c)]{BSB05} Bahcall, J.~N., Serenelli, A.~M., \& Basu, S.\ 2005c, \apjl, 621, L85 
\bibitem[Bahcall et al.(2001)]{Bahcall2001} Bahcall, J.~N., Pinsonneault, M.~H., \& Basu, S.\ 2001, \apj, 555, 990 
\bibitem[Bahcall \& Pinsonneault(2004)]{BP04} Bahcall, J. N., \& Pinsonneault, M. H. 2004, Phys. Rev. Lett., 92, 121301
\bibitem[Bahcall et al.(2004)]{BSP04} Bahcall, J. N., Serenelli, A. M., \& Pinsonneault, M. H. 2004, \apj, 614, 464
\bibitem[Basu(1998)]{Basu1998} Basu, S. 1998, MNRAS, 298, 719
\bibitem[Basu \& Antia(2004)]{BA2004} Basu, S., \& Antia, H. M. 2004, ApJ, 606, L85
\bibitem[Basu \& Antia(2008)]{BA08} Basu, S., \& Antia, H.~M.\ 2008, \physrep, 457, 217 
\bibitem[Basu et al.(2000)]{Basu2000} Basu, S., Pinsonneault, M.~H., \& Bahcall, J.~N.\ 2000, \apj, 529, 1084 
\bibitem[Brown et al.(2007)]{s34b} Brown, T.~A.~D., Bordeanu, C., Snover, K.~A., Storm, D.~W., Melconian, D., Sallaska, A.~L., Sjue, S.~K.~L., \& Triambak, S.\ 2007, \prc, 76, 055801 
\bibitem[Caffau et al.(2008)]{caffau08} Caffau, E., Ludwig, H.-G., Steffen, M., Ayres, T.~R., Bonifacio, P., Cayrel, R., Freytag, B., \& Plez, B.\ 2008, \aap, 488, 1031 
\bibitem[Caffau et al.(2010)]{caffau10} Caffau, E., Ludwig, H.-G., Steffen, M., Freytag, B., \& Bonifacio, P.\ 2010, arXiv:1003.1190 
\bibitem[Chaplin et al.(2007)]{Chaplin07} Chaplin, W.~J., Serenelli, A.~M., Basu, S., Elsworth, Y., New, R., \& Verner, G.~A.\ 2007, \apj, 670, 872 
\bibitem[Confortola et al.(2007)]{s34a} Confortola, F., et al.\ 2007, \prc, 75, 065803 
\bibitem[Delahaye \& Pinsonneault(2006)]{DP06} Delahaye, F., \& Pinsonneault, M.~H.\ 2006, \apj, 649, 529 
\bibitem[Di Mauro et al.(2002)]{DiMauro2002} Di Mauro, M. P., Christensen-Dalsgaard, J., Rabello-Soares, M. C., \& Basu, S. 2002, 
\aap, 384, 666
\bibitem[Drake \& Testa(2005)]{drake05} Drake, J.~J., \& Testa, P.\ 2005, \nat, 436, 525 
\bibitem[Drake \& Ercolano(2007)]{drake07} Drake, J.~J., \& Ercolano, B.\ 2007, \apjl, 665, L175 
\bibitem[Ferguson et al.(2005)]{Ferguson2005} Ferguson, J.~W., Alexander, D.~R., Allard, F., Barman, T., Bodnarik, J.~G., Hauschildt, P.~H., Heffner-Wong, A., \& Tamanai, A.\ 2005, \apj, 623, 585 
\bibitem[Grevesse \& Noels(1993)]{gn93} Grevesse, N., \& Noels, A.\ 1993, Physica Scripta Volume T, 47, 133 
\bibitem[Grevesse \& Sauval(1998)]{gs98} Grevesse, N., \& Sauval, A. J. 1998, Space Sci. Rev., 85, 161
\bibitem[Guzik et al.(2005)]{Guzik05} Guzik, J. A., Watson, L. W., \& Cox, A. N. 2005, ApJ 627, 1049
\bibitem[Junghans et al.(2003)]{s17} Junghans, A.~R., et al.\ 2003, \prc, 68, 065803 
\bibitem[Koesterke et al.(2008)]{Koesterke08} Koesterke, L.,Allende Prieto, C., \& Lambert, D.~L.\ 2008, \apj, 680, 764 
\bibitem[Lodders et al.(2009)]{Lodders2009} Lodders, K. Palme, H. Gail. H.P. 2009, Landolt Boernstein, New Ser. in press
\bibitem[Marta et al.(2008)]{s114} Marta, M., et al.\ 2008, \prc, 78, 022802 
\bibitem[Montalb\'an et al.(2004)]{Montalban04} Montalban, J., Miglio, A., Noels, A., Grevesse, N., \& di Mauro, M. P. 2004, SOHO 14, Helio- and Astroseismology: Towards a Golden Future, 559, 574 
\bibitem[Pinsonneault \& Delahaye(2009)]{PD09} Pinsonneault, M.~H., \& Delahaye, F.\ 2009, \apj, 704, 1174 
\bibitem[Pinsonneault(1997)]{Pinsonneault1997} Pinsonneault, M.\ 1997, \araa, 35, 557 
\bibitem[Richard et al.(1998)]{Richard1998} Richard, O., Dziembowski, W.~A., Sienkiewicz, R., \& Goode, P.~R.\ 1998, \aap, 338, 756 
\bibitem[Richard et al.(1996)]{Richard1996} Richard, O., Vauclair, S., Charbonnel, C., \& Dziembowski, W.~A.\ 1996, \aap, 312, 1000 
\bibitem[Rogers \& Nayfonov(2002)]{eos2006} Rogers, F.~J., \& Nayfonov, A.\ 2002, \apj, 576, 1064 
\bibitem[Schmelz et al.(2005)]{Schmelz05} Schmelz, J.T., Nasraoui, K.; Roames, J.K.; Lippner, L.A. \& Garst, J.W. 2005 \apjl, 634, L197
\bibitem[Seaton et al.(1994)]{SYMP1994} Seaton, M.~J., Yan, Y., Mihalas, D., \& Pradhan, A.~K.\ 1994, \mnras, 266, 805 
\bibitem[Seaton \& Badnell(2004)]{Seaton04} Seaton, M. J., \& Badnell, N. R. 2004, MNRAS, 354, 457
\bibitem[Turck-Chi{\`e}ze et al.(2004)]{Turck04} Turck-Chi{\`e}ze, S., et al. 2004, Phys. Rev. Lett., 93, 211102 Widing, K. G. 1997, ApJ, 480, 400
\bibitem[Young(2005)]{Young05} Young, P.R. 2005, \aap, 439, 361 
\bibitem[review by Zeippen(1995)]{CJZ1995} Zeippen, C.~J.\ 1995, Physica Scripta Volume T, 58, 43 
\end{thebibliography}
\end{document}